\pdfoutput=1

\documentclass[11pt]{article}

\usepackage{emnlp2021}

\usepackage{times}
\usepackage{latexsym}

\usepackage[T1]{fontenc}
\usepackage{flushend}
\usepackage[a-1b]{pdfx}   
\usepackage{balance}

\usepackage{color}
\usepackage{tcolorbox}
\usepackage[labelfont=bf]{caption}
\usepackage{subcaption}
\usepackage{hyperref}
\usepackage{makecell}
\usepackage{fixltx2e}
\usepackage{rotating,multirow}
\usepackage{etoolbox,xspace}
\usepackage{algorithmicx}
\usepackage{algorithm}
\usepackage{fontawesome}
\usepackage{algpseudocode}
\usepackage{wrapfig}
\usepackage{scalerel,amssymb}
\usepackage{booktabs}
\usepackage{multirow}
\usepackage{siunitx}
\usepackage{amsmath}

\algtext*{EndWhile}
\algtext*{EndIf}
\algtext*{EndFunction}
\algtext*{EndFor}
\algrenewcommand\algorithmicrequire{\textbf{Input:}}
\algrenewcommand\algorithmicensure{\textbf{Output:}}

\usepackage{amsthm}
\usepackage{booktabs}
\usepackage{colortbl}
\usepackage{xcolor}

\usepackage{tabularx}
\usepackage{ragged2e}  
\usepackage{booktabs}  
\usepackage{textcomp}
\usepackage{url}
\usepackage{comment}
\usepackage{enumitem}
\usepackage[simplified]{pgf-umlcd}
\usepackage{tikz}

\usepackage{tabularx}                       
\newcolumntype{C}{>{\centering\arraybackslash}X} 

\newcolumntype{Y}{>{\RaggedRight\arraybackslash}X}

\newtcolorbox{defbox}{
  colback=orange!10,
  colframe=orange!20,
  arc=2mm, 
  fonttitle=\bfseries,
  boxrule=0mm,
  boxsep=1mm,
  left=0mm,
  right=0mm,
  top=0mm,
  bottom=0mm
}

\newcounter{example}
\renewcommand{\theexample}{\arabic{example}}

\newtcolorbox{examplebox}{
  colback=blue!10,
  colframe=blue!20,
  arc=2mm, 
  fonttitle=\bfseries,
  boxrule=0mm,
  boxsep=1mm,
  left=0mm,
  right=0mm,
  top=0mm,
  bottom=0mm
}

\newcommand{\squishlist}{
 \begin{list}{$\bullet$}
  { \setlength{\itemsep}{0pt}
     \setlength{\parsep}{1pt}
     \setlength{\topsep}{1pt}
     \setlength{\partopsep}{0pt}
     \setlength{\leftmargin}{1em}
     \setlength{\labelwidth}{1em}
     \setlength{\labelsep}{0.5em} } }

\newcommand{\squishend}{
  \end{list}
}

\definecolor{americanrose}{rgb}{1.0, 0.01, 0.24}
\definecolor{airforceblue}{rgb}{0.36, 0.54, 0.66}
\definecolor{ao(english)}{rgb}{0.0, 0.5, 0.0}
\definecolor{ao}{rgb}{0.0, 0.0, 1.0}
\definecolor{darkgreen}{rgb}{0.0, 0.5, 0.0}

\newcommand{\eat}[1]{}

\usepackage{xspace}

\usepackage[utf8]{inputenc}

\usepackage{microtype}

\title{An Adversary-Resistant Multi-Agent LLM System via Credibility Scoring}

\author{Sana Ebrahimi \\
    University of Illinois Chicago \\
  \texttt{sebrah7@uic.edu} \\\And
  Mohsen Dehghankar \\
  University of Illinois Chicago\\
  \texttt{mdehgh2@uic.edu} \\\And
  Abolfazl Asudeh \\
  University of Illinois Chicago\\
  \texttt{asudeh@uic.edu}\\}

\begin{document}
\maketitle
\begin{abstract}
    While multi-agent LLM systems show strong capabilities in various domains, they are highly vulnerable to adversarial and low-performing agents.
    To resolve this issue, in this paper, we introduce a general and adversary-resistant multi-agent LLM framework based on credibility scoring.
    We model the collaborative query-answering process as an iterative game, where the agents communicate and contribute to a final system output.
    Our system associates a credibility score that is used when aggregating the team outputs. The credibility scores are learned gradually based on the past contributions of each agent in query answering.
    Our experiments across multiple tasks and settings demonstrate our system’s effectiveness in mitigating adversarial influence and enhancing the resilience of multi-agent cooperation, even in the adversary-majority settings.    
\end{abstract}

\section{Introduction}\label{sec:intro}

Multi-agent LLM systems have risen as a powerful paradigm, exemplified by frameworks such as CAMEL, AutoGen, and MetaGPT \cite{autogen, metagpt, camel}, demonstrating promising performance in crucial domains, including coding, mathematical problem-solving, and collaborative decision-making.

Despite their advancements, the performance of multi-agent LLM systems is highly sensitive to adversarial and low-performing agents. 
Particularly, a subset of compromised team members with adversarial behavior can corrupt the system’s collective output. The susceptibility of LLM agents to persuasive inputs further amplifies this risk, potentially leading to incorrect group consensus. 
Although prior studies have highlighted this vulnerability \cite{psysafe, multiagent-debate-attack, MALLM-survey-2}, existing solutions are predominantly limited to specific, predefined architectures. These approaches and the related work are further discussed in Appendix~\ref{app:related}.

To the best of our knowledge, the literature lacks a general framework that enables users to design {\em robust multi-agent systems resilient to adversarial influence} while minimizing the impact of such attacks without the need to eliminate an agent.

In this paper, we fill this research gap by proposing an {\em adversary-resistant} multi-agent LLM system based on {\em credibility scoring.}

Specifically, we model the query-answering process as an iterative cooperative game, where a team of agents is formed to find the answer to a given query. The team members may have different roles and communicate based on the team's topology to finalize their individual answers, which are then aggregated into the system's answer to the query.

Instead of equally trusting all agents, our system follows a credibility-score aware aggregation strategy that weighs each agent's individual output proportional to their trustworthiness.
The credibility scores reflect the collective performance of each agent in answering the previous queries and are learned on the fly during the lifetime of the system.

For each query, the team receives a reward (or gets penalized) based on the quality of the generated output.
In order to fairly distribute the reward among the team members, we introduce the {\em contribution scores}, with larger values reflecting a larger impact of an agent in the generated output.
We propose two approaches based on Shapley values and LLM-as-Judge for measuring the contribution scores.
At the end of each round, the credibility scores are updated by distributing the reward to the agents proportional to their contribution. 

Our system has a unique ability to tolerate adversary-majority settings, a more extreme case than the typically considered settings that assume the adversaries are in the minority. We emphasize a critical yet under-explored challenge: when adversaries constitute more than 50\% of the agents, honest agents must either exert disproportionate influence or possess superior capabilities to avoid being outvoted or manipulated. 

Our approach is applicable across different team structures and integration mechanisms for existing methods. It empowers users to 
minimize the impact of low-performing and malicious agents within the teams with various formations and communications topologies. By leveraging this adaptability, our method enhances the resilience of multi-agent systems, ensuring more robust and reliable cooperation of the agents. 
We conduct comprehensive experiments on various tasks, benchmark datasets, and settings to evaluate our system.
Our experiment results verify the effectiveness of credibility scoring, 
demonstrating the ability of our system in detecting and minimizing the effect of the adversary agents, even for the adversary-majority settings.
\vspace{-0.5mm}
\paragraph{Paper Organization:} 
We first introduce the concepts and provide an overview of our system in Section~\ref{sec:overview}. Next in Section~\ref{sec:team}, we discuss the composition details of a team of agents, followed by the explanation of the credibility-score aware aggregation of the team outputs in Section~\ref{sec:agg}. 
We then conclude our technical discussions in Section~\ref{sec:tech-details} by explaining how the credibility scores are gradually learned in our system.
The experimental evaluations are provided in Section~\ref{sec:experiments}, followed by the concluding remarks and a discussion of our system's limitations in Sections~\ref{sec:conclusion} and~\ref{sec:limitations}.
\begin{figure}[tb]                
  \centering
  \includegraphics[width=.49\textwidth]{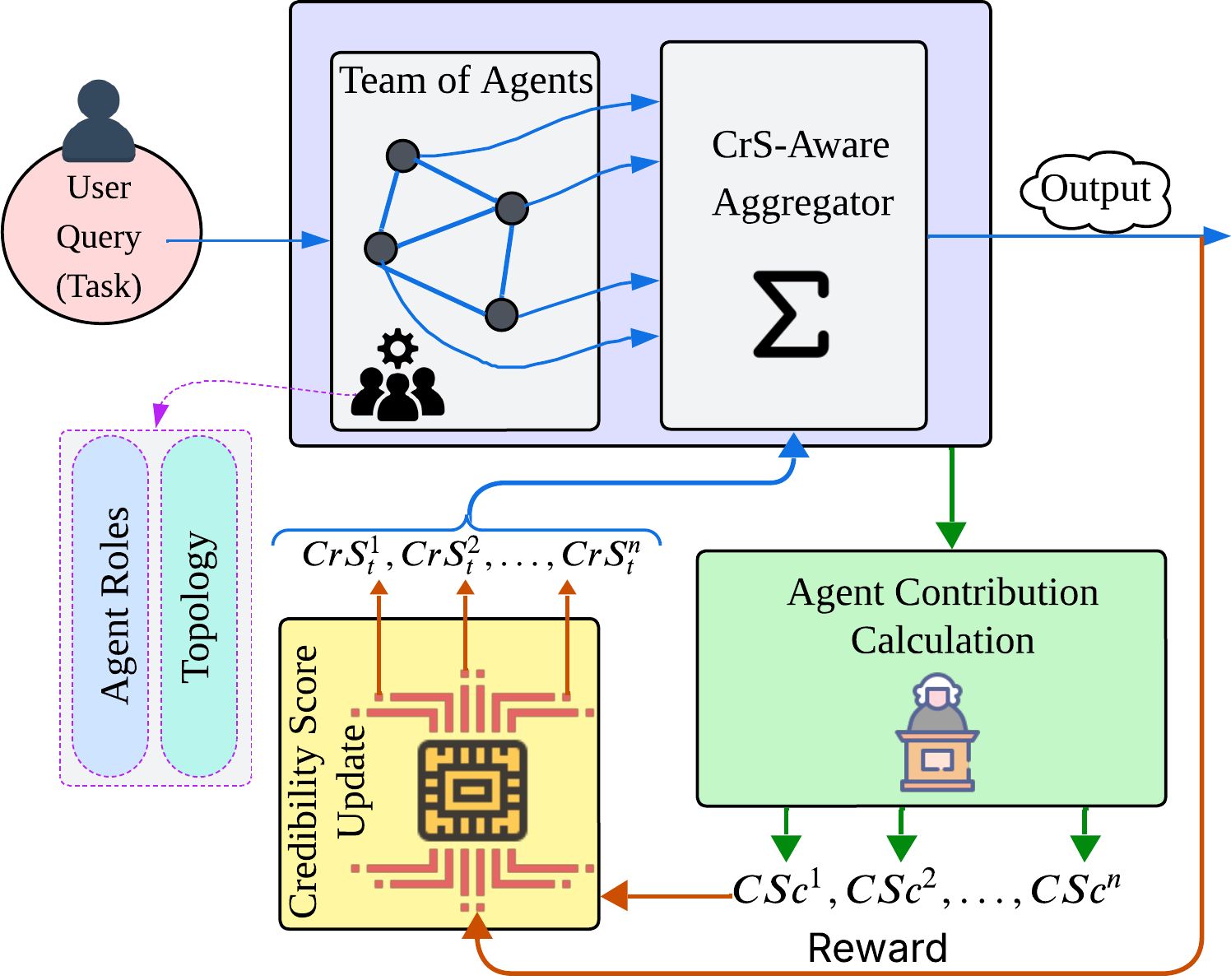}
  \vspace{-5mm}
  \caption{System architecture.}
  \label{fig:sys-arch}
  \vspace{-5mm}
\end{figure}

\section{System Overview}\label{sec:overview}

We consider a system, with the architecture shown in Figure~\ref{fig:sys-arch}, that uses a universe $\mathcal{A}$ of LLM agents for answering user queries specified in the form of natural language instructions, known as {\em prompts}.

The answer to each query $q$ is generated by a {\bf team of agents} $ A= \{a_1,\cdots,a_N\}\subseteq \mathcal{A}$.

We model this system as an iterative cooperative game, where at each iteration $t$, a team $A_t$ is formed based on a specific {\bf topology} 
that specifies the communication rules, while the agents may have various {\bf roles} in the team. 
The team members collaborate, and each agent, in the end, generates an output. 
In Section~\ref{sec:team}, we shall further discuss the structure of the teams of agents.

Our system then {\bf aggregates} the individual outputs to generate the final output for the user query, using the {\bf Credibility Score}: 
we allocate each agent $a_j$ with a credibility score $CrS^{(j)}\in [0,1]$, a numerical value that reflects the collective reliability of $a_j$ over the previous iterations.
The credibility scores of the agents are gradually ``learned'' during the life time of the system (see Section~\ref{sec:tech-details}).

Introducing the credibility scores gives our system the unique feature to be able {\em to tolerate and detect} {\bf malicious agents} with adversarial behaviors. While {\em faithful} agents pursue a correct solution, the {\em adversarial} agents deliberately attempt to mislead or derail the group to generate wrong answers.
We extensively evaluate the robustness of our system in our experiments (Section~\ref{sec:experiments}). 

For each generated answer (final output) $o_t$ for the query $q_t$, we consider a {\bf reward} $r_t\in [-1,1]$, specified based on the ``quality'' of $o_t$ as an answer for $q_t$.
Specifically, a negative value of $r_t$ {\em penalizes} the team $A_t$ for generating a misleading result, while a positive $r_t$ rewards the team for generating a good answer. 

We view $o_t$ as the outcome of the team's {\em collective effort} and {\em distribute the reward} among agents in proportion to their, {\em ``contribution''} to generating $o_t$.
We introduce the {\bf Contribution Score} (CSc) to measure each team member's contribution. 

Finally, we update the credibility score of each team member $a_i\in A_t$, using a learning step, based on the amount of the reward $a_i$ collected by collaborating in answering the query $q_t$.
In Section~\ref{sec:tech-details}, we shall provide the technical details of this process.





\section{Team of Agents}\label{sec:team}

In this section, we explain the key components in the formation of a team of agents, including the topology and the gent roles.
\vspace{-0.5mm}
\paragraph{Agent Roles.}
In a multi-agent LLM system all team members may be assigned to the same task~\cite{DyLAN, div-think-llms}, or they may have different roles aligned with their specific expertise or subtasks \cite{CoA, MA-LLM-topo}. 
Another important consideration is the agents' adaptability: whether they can learn, adapt, or modify their strategies over time by updating internal parameters. 
These aspects have been explored in varying degrees across existing research. For instance, Alfonso et al.~\cite{multiagent-debate-attack} demonstrated that models could be influenced to alter their behavior in ways that ultimately degrade overall system performance. Such interference may occur through direct manipulation of agents' individual contributions or deceptive communication tactics \cite{multiagent-debate-attack}. Further details about incentives and adversarial behavior of LLM agents is discussed in Appendix~\ref{app:behavior}.

Therefore, establishing robust mechanisms to mitigate adversarial threats is essential to maintaining the integrity and reliability of multi-agent collaborations.
A major benefit of our systems is the {\bf robustness against adversarial agents}.
Specifically, allocating the agents with credibility scores, our system gradually penalizes the agents with adversarial behaviors (see Section~\ref{sec:tech-details}).
In Section~\ref{sec:experiments}, we demonstrate that our system can tolerate {\em even more than half} of the agents being adversary.

\paragraph{Communication Structure (Topology).}
The topology of a multi-agent system defines the arrangement and interconnections among agents, effectively determining which agents can directly communicate. This structure can be conceptualized as a graph, where each node represents an agent, and each edge represents a direct communication link between two agents. 
Previous research has investigated various topological arrangements from (a) {\em no connection} to (b) {\em fully-connected} structures, including (c) {\em chain}, (d) {\em ring}, (e) {\em hierarchical}, and (f) {\em randomly-connected} networks~\cite{MoA, resilience, MA-LLM-topo, DyLAN}. 
The choice of topology significantly impacts both scalability and robustness of the multi-agent system. For example, fully connected topologies facilitate rapid consensus due to their direct communication paths, yet they exhibit vulnerability when faced with adversarial agents or limited network resources~\cite{multiagent-debate-attack}. Conversely, sparse topologies, such as ring or chain structures, lower communication overhead but might be more susceptible to localized adversarial influence, potentially compromising subsets of agents \cite{multi-agent-sys}. 

Our system is flexible to the choice of communication structures:
we assume each agent first drafts a candidate solution (\emph{local inference}). Then, during the 
{\em peer interaction} phase, the agents optionally exchange information according to the \emph{communication graph} prescribed by the topology.

\section{CrS-Aware Aggregation}\label{sec:agg}

As illustrated in Figure~\ref{fig:sys-arch}, after peer interactions, each agent $a_i\in A_t$ generates an output.
Existing coordination mechanisms (discussed in Appendix~\ref{app:coord}) integrate these outputs into an answer to the user query, following the strategies such as majority voting.

Building on top of the existing aggregation schemes, our system adds {\em credibility-scores} (CrS) to make the final output more reliable and robust against adversaries and low-performing agents.

Formally, the credibility score $CrS^{(j)}\in [0,1]$
of an agent $a_j\in \mathcal{A}$ is a {non-negative} number that reflects how reliable the system views the agent $a_i$ according to its performance in the previous query answering rounds. 

The credibility scores can be used in various coordination mechanism by replacing the unweighted aggregation with the weighted aggregation using the CrS scores.
Without loss of generality, in the following, we illustrate their integration into two integration mechanisms:





\vspace{1mm}
\noindent{\em (a) centroid-based aggregation:} \label{sec:agg-requal}
\cite{requal} proposes an aggregation strategy that first finds the centroid of the generated outputs in the embedding space, and then returns the closest answer to the centroid as the final output (see Appendix~\ref{app:related} for more details).
We use the CrS scores to find the {\em CrS-aware centroid} $\vec{x}^+$ as the weighted average of the generated outputs:

\vspace{-5mm}
\begin{equation}
\vec{x}^{+}=\frac{1}{N} \sum_{i : a_i\in A_t} {CrS}_{t-1}^{(i)} \vec{v}\left( O(a_i,q_t)\right)
\end{equation}

\vspace{-2mm}\noindent 
Where $O(a_i,q_t)$ is the output of agent $a_i$ for $q_t$, $\vec{v}(o)$ is the embedding of an output $o$, and ${CrS}_{t-1}^{(i)}$ is the current credibility score of $a_i$. 

\vspace{1mm}
\noindent{\em (b) LLM-assisted aggregation:}
Instead of using a specific formula for aggregation, one can use an LLM for this step, where in addition to the outputs $o_1,\cdots,o_N$, the {\em CrS scores} of the participating agents are sent to a {\bf Credibility‑aware Coordinator LLM}, which we trust.
The coordinator then aggregates the individual outputs and generates the final output while considering the CrS scores.


\section{Learning Credibility Scores On-The-Fly}\label{sec:tech-details}
 
Our system learns the credibility scores of the agents on the fly based on their performance in answering previous queries $\{q_1,\cdots, q_{t-1}\}$.

Initially, assuming there are no prior information about the reliability of the agents, all credibility scores are set to a default value (e.g., 0.5).
Then, at the end of each round $t$, the system computes a {\bf contribution score} $CSc^{(i)}$ for each of the team members $a_i\in A_t$.

Depending on the quality of the generated answer $o_t$ for the query $q_t$, the team is rewarded with a value $r_t\in[-1,1]$. The contribution scores and the reward value are then used for updating the credibility scores. The computation of the reward values is discussed in Appendix~\ref{app:reward}.

In the following, we first discuss the computation of the contribution scores, and then explain how the credibility scores are updated.

\subsection{Calculating the agent contributions}\label{sec:tech-details-Csc}

Given a query $q_t$, we model the process of generating the output $o_t$ as a game, where the team members collaboratively obtain the reward $r_t$.
Since the team members may have different impacts in the generation of the output, their share of the reward should be proportional to their contribution.

We propose the following approaches for computing the contribution scores (CSc):

\vspace{1mm}
\noindent{\em (i) Shapley Values for CSc computation:}
Our first approach for computing the contribution scores is based on {\em Shapley values} -- the popular concept in Game Theory for fairly distributing the reward among a team of players who have collaborated~\cite{shapley1951notes}.
Specifically, we consider Shapley values for {\em no-communication} topologies and when the aggregation strategy {\em is not} LLM-assisted.

Let $O=\{o_1,\cdots,o_N\}$ be the set of individual responses generated by a agents $A_t = \{a_1,\cdots,a_N\}$.
Let $\Sigma(S)$ be the final output generated by aggregating the responses of a subset of responses $S\subseteq O$.
Also, let $R(o_t)$ be the reward allocated based on the quality of $o_t$ as the answer of the query $q_t$.
The contribution score of the agent $a_i$ is then computed using the following equation:
\begin{align*}
 CSc^{(i)} =
\sum_{S\subseteq R\setminus{o_i}} \tfrac{|S|!(N-|S|-1)!}{N!}\Big(R&(\Sigma\big(S\cup{x_n})\big)\\
&-R\big(\Sigma(S)\big)\Big)
\end{align*}

\vspace{-3mm}
\noindent{\em (ii) LLM-as-Judge for CSc computation:}
Despite their advantages such as theoretical guarantees, {\em it is \#P-hard} to compute Shapley values. As a result, computing the CSc values based on Shapley values require a {\em combinatorial} number of reward value computations for the aggregated outputs generated by each subset of $(A_t\backslash a_i)$.
This makes it practically infeasible to compute the contribution scores for the following cases.
(A) When the team members communicate, their output may be impacted by the composition of the team. As a result, for each subset $S\subseteq A_t$, one would need to form a new team and observe new outputs.
(B) When the aggregation of reward value computation is LLM-assisted, an LLM query would be needed for each subset $S\subseteq A_t$ to compute the reward.

Therefore, we instead use an LLM Judge to compute the contribution scores in such settings.
Specifically, given a query $q_t$,
we send the final answer $o_t$, the dialogue log, and the agent outputs to the LLM Judge, and ask the judge to quantify the contribution of each agent in the generation of the final output $o_t$.
The judge can analyze the message-passing log and observe which agents changed their response after the communication. The Agents never see these numbers to prevent strategic gaming.

\subsection{Updating the CrS values}  

Once the contribution scores are computed, the credibility score of each agent gets updated by distributing the reward $r_t$ among the agents proportional to their contribution. 
Specifically, using a learning rate $\eta$, the credibility scores are updated using Equation~\ref{eq:crs}.
\begin{equation}
\operatorname{CrS}^{(i)}_{t}=\operatorname{CrS}^{(i)}_{t-1}\bigl(1 + \eta . \operatorname{CSc}^{(i)}.r_t\bigr)
\label{eq:crs}
\end{equation}

Before concluding this section, we would like to remind that our scoring mechanism for computing CSc and CrS values is orthogonal to the team formation details including the agent roles and communication structure, making it easy to operate on top of the existing multi‑agent toolkits such as \textsc{AutoGen} and \textsc{CAMEL}.  Source code and prompts are provided in the supplementary material.
\section{Experiment Results}\label{sec:experiments}


\begin{table*}[ht]
\centering
\small
\caption{Accuracy results for multi-agent LLMs using LLaMA 3.2 3B, Mistral 7B, and Qwen2.5 7B. \textit{CrS} indicates use of the Credibility Scoring mechanism, and the accuracy gain over naive coordination is denoted by $\Delta$.}
\begin{tabular}{|l|c|cc|cc|cc|cc|}
\toprule
\textbf{Backbone Model} & \textbf{Architecture} & \multicolumn{2}{c|}{\textbf{GSM8K}} & \multicolumn{2}{c|}{\textbf{MMLU-MS}} & \multicolumn{2}{c|}{\textbf{MATH}} & \multicolumn{2}{c|}{\textbf{Research QA}} \\
\cmidrule(r){3-4} \cmidrule(r){5-6} \cmidrule(r){7-8} \cmidrule(r){9-10}
&& CrS & $\Delta$ & CrS  & $\Delta$ & CrS & $\Delta$ & CrS & $\Delta$ \\
\midrule
\multirow{2}{*}{LLaMA 3.2(3B)} 
& SIA & 47.5 & \textcolor{darkgreen}{$+8\%$} & 35.5 & \textcolor{darkgreen}{$+15\%$} & 40.0 & \textcolor{darkgreen}{$+7\%$} & 52.0 & \textcolor{darkgreen}{$+51\%$} \\
& CrS‑ordered Chain & 43.0  & \textcolor{darkgreen}{$+20\%$} & 44.0 & \textcolor{darkgreen}{$+16\%$} & 32.0 & \textcolor{darkgreen}{$+15\%$} & 84.0 & \textcolor{darkgreen}{$+20\%$} \\
\midrule
\multirow{2}{*}{Mistral(7B)} 
&SIA& 12.0 & \textcolor{darkgreen}{$+6\%$} & 21.0 & \textcolor{darkgreen}{$+9\%$} & 11.5 & \textcolor{darkgreen}{$+5.5\%$} & 86.0 & \textcolor{darkgreen}{$+14\%$} \\
& CrS‑ordered Chain& 13.0 & \textcolor{darkgreen}{$+11\%$} & 32.0 & \textcolor{darkgreen}{$+6\%$} & 08.0 & \textcolor{darkgreen}{$+6\%$} & 77.0 & \textcolor{red}{$-7\%$} \\
\midrule
\multirow{2}{*}{Qwen2.5(7B)} 
& SIA& 75.5 & \textcolor{darkgreen}{$+10.5\%$} & 43.0 & \textcolor{darkgreen}{$+25.5\%$} & 65.0 & - & 59.0 & \textcolor{darkgreen}{$+17\%$} \\
& CrS‑ordered Chain& 60.0 & \textcolor{darkgreen}{$+10\%$} & 52.0 & \textcolor{darkgreen}{$+10\%$} & 59.8 & \textcolor{darkgreen}{$+9\%$} & 90.0 & \textcolor{darkgreen}{$+5\%$} \\
\bottomrule
\end{tabular}
\label{tab:progress-results}
\end{table*}

\eat{
HumanEval dataset and ResearchQA should be discussed separately. The results show that for these type of datasets, a structured architecture shows a better results.
}
\subsection{Experiments Setting}

\paragraph{Backbone Models \& Datasets.} 
We deploy three lightweight open-source LLMs; \textbf{Llama3.2(3B)}~\cite{llama3.2}, \textbf{Mistral(7B)}~\cite{mistral7b} and \textbf{Qwen2.5(7B)}~\cite{qwen2.5} as both agents and coordinator, allowing cost-efficient scaling while testing models that remain susceptible to adversarial noise. A stronger GPT-4o mini~\cite{gpt4o-mini} acts as an external judge, evaluating and scoring the team's final answers. We evaluate our framework on five benchmarks: \textbf{ MMLU-MS}~\cite{MMLU} (Math and Statistics), \textbf{MATH}~\cite{math}, \textbf{GSM8K}~\cite{gsm8k}(open‑ended mathematical reasoning), \textbf{HumanEval}~\cite{humaneval}(code completion), and \textbf{Research Questions}~\cite{researchQA}(non‑factoid, search‑style questions requiring contextual judgment). Together, these benchmarks evaluate the system’s robustness, mathematical and factual reasoning skills, and coding competence. Comprehensive information on model selection, data preprocessing, and evaluation procedures is available in Appendix~\ref{app:exp-details}.

\paragraph{Compared Methods.} For comparison, we implement three baseline methods: single-agent response generation, naive coordination, majority voting, and similarity-based ensemble  approaches. In the similarity‑based ensemble of \cite{more-agents}, each answer is compared with every other answer, and the one with the largest total pairwise similarity is selected as the final response. In the single-agent baseline scenario, the final team response is selected from one of the faithful agents, randomly designated as the coordinator, after completing multi-agent communication. All reported experimental results represent the final output produced after comprehensive internal communication among agents. Finally, the naive coordination uses the same LLM coordinator, but it produces the final answer without receiving the agents’ credibility. 

\eat{Our preliminary experiments suggested that achieving clear and interpretable outcomes necessitates a judge substantially more capable than the evaluated agents. Accordingly, we utilize GPT-4o-mini, accessed through API calls, as the evaluation judge. Occasionally, communication histories may exceed token limits. To mitigate this issue, we employ LLMLingua~\cite{llmlingua} to compress the prompt provided to the judge as necessary. Importantly, none of the participating agents, coordinator, or judge models have undergone specific pre-training for the evaluation tasks they perform.}
\begin{figure}[t]              
  \centering
  \begin{subfigure}[b]{0.4\textwidth}
    \centering
    \includegraphics[width=\linewidth]{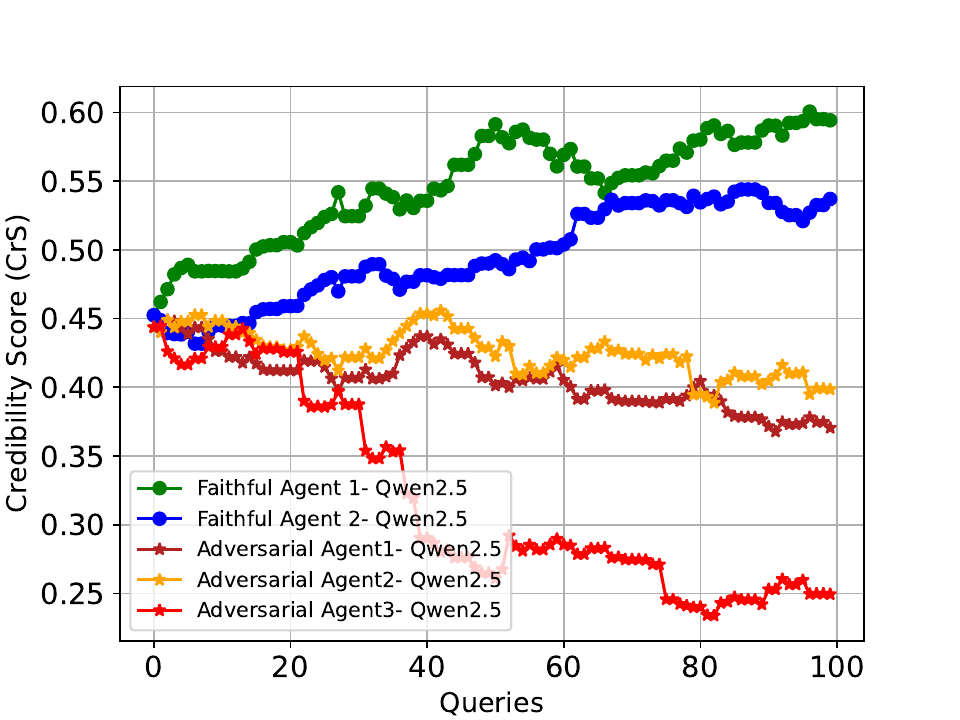}
    \caption{Qwen 2.5 agents on GSM8K.}
    \label{fig:judge-weight-a}
  \end{subfigure}%
  \vspace{-1mm}
  \hfill
  \begin{subfigure}[b]{0.4\textwidth}
    \centering
    \includegraphics[width=\linewidth]{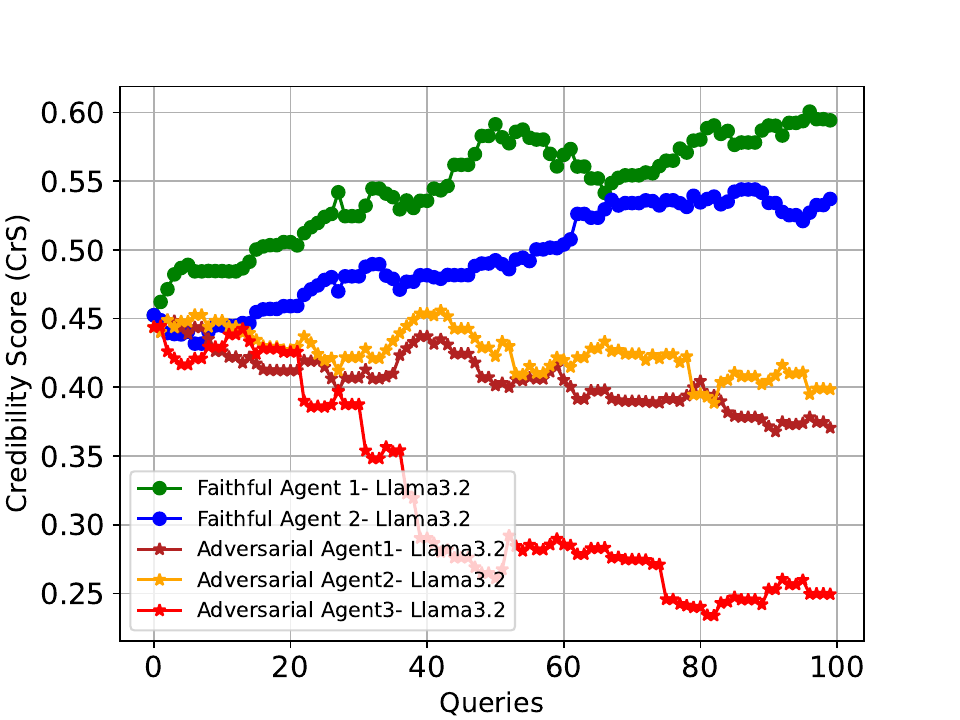}
    \caption{LLaMA 3.2 agents on ResearchQA.}
    \label{fig:judge-weight-b}
  \end{subfigure}
  \vspace{-2mm}  
  \caption{CrS convergence for an adversary-dominated team with 3 adversarial and 2 faithful agents.}
  \label{fig:judge-weight}
  \vspace{-4mm}
\end{figure}

\subsection{Collaboration Setup}

We run our primary experiments with five agents. Two faithful and three adversarial ones, that inject subtle errors, are prompted using similar prompt template across tasks. We evaluate our method across three communication topologies\footnote{
The implementation details are provided in Appendix~\ref{app:exp-details}.}:

\textbf{Stochastic Interaction Architecture (SIA).} For each question, six undirected links are sampled at random from the ${5\choose 2}$ possible pairs, yielding diverse topologies such as trees, rings, etc. Agents may review or maintain their own answers after reading the messages from their neighbors.

\textbf{Standalone Agent Architecture (SAA).} Each agent responds independently without any peer interaction. Finally a centroid‑based aggregation \cite{requal} is used to select the team’s answer by choosing the nearest response to the centroid of all outputs as discussed in Section~\ref{sec:agg-requal}.

\textbf{Credibility-ordered Chain.} We additionally evaluate a CrS‑ordered chain topology. In this setting, agents are arranged by their current credibility scores and exchange messages only with neighbors.



\subsection{Insights from Experimental Observations}

\begin{figure*}[t]               
  \centering
  \subcaptionbox{MMLU-Llama3.2\label{fig:plotA}}[0.30\textwidth]{%
    \includegraphics[width=\linewidth]{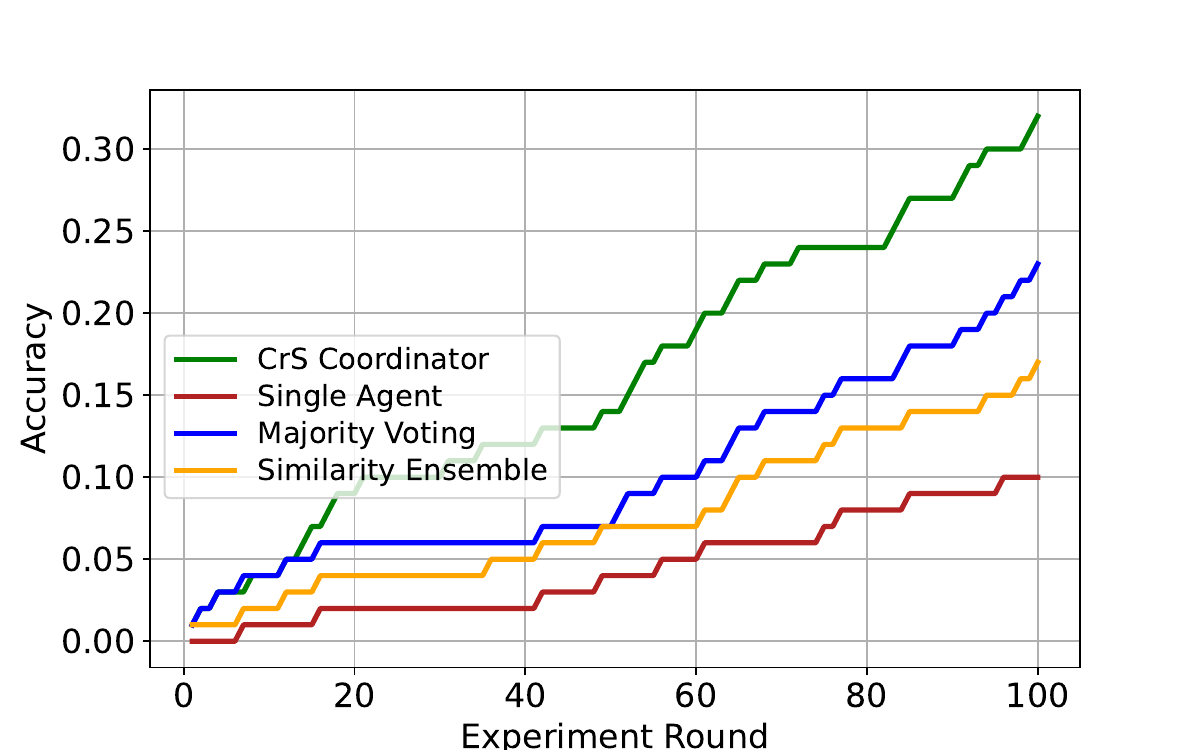}}
  \hfill
  \subcaptionbox{MMLU-Mistral\label{fig:plotB}}[0.30\textwidth]{%
    \includegraphics[width=\linewidth]{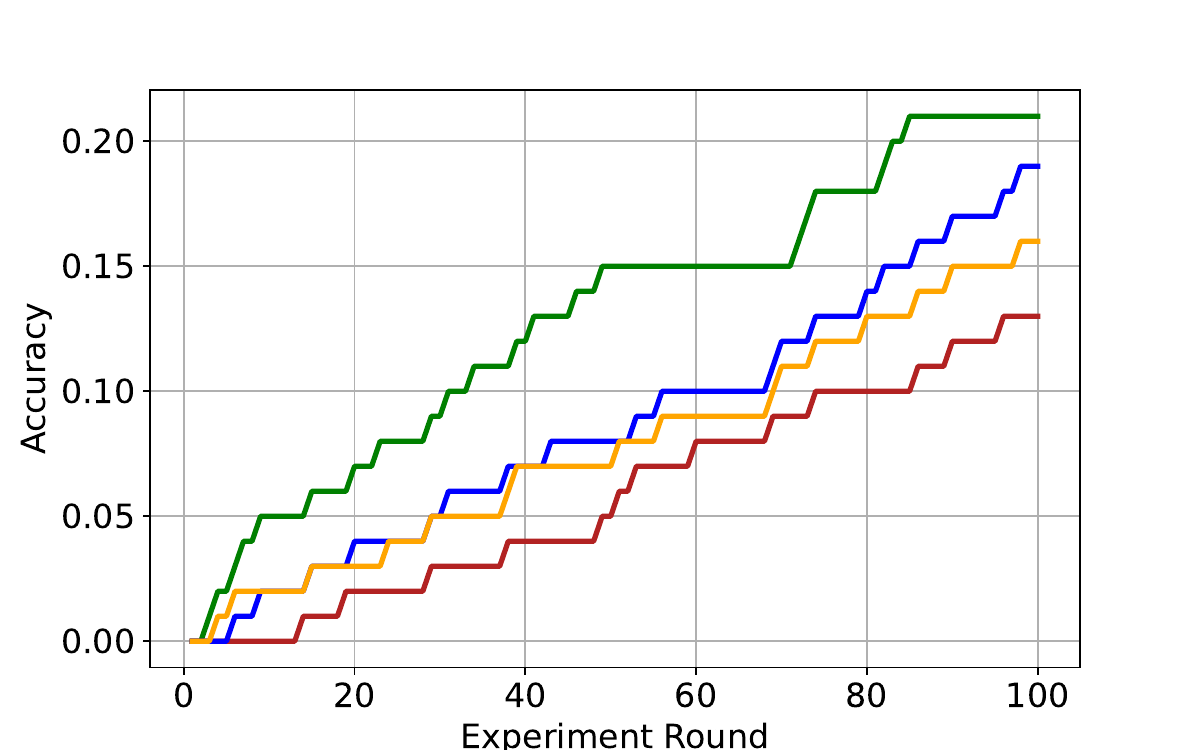}}
  \hfill
  \subcaptionbox{MMLU-Qwen2.5\label{fig:plotC}}[0.30\textwidth]{%
    \includegraphics[width=\linewidth]{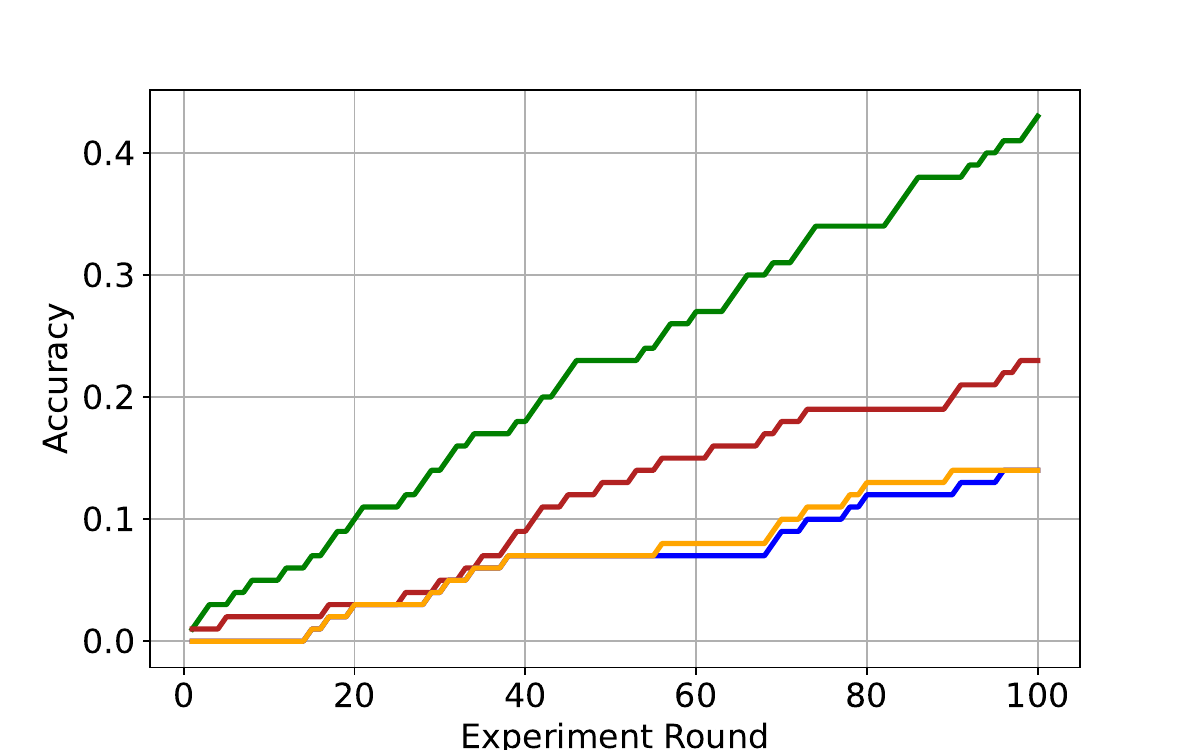}}

  \vspace{-1mm}

  \subcaptionbox{GSM8K-Llama3.2\label{fig:plotD}}[0.30\textwidth]{%
    \includegraphics[width=\linewidth]{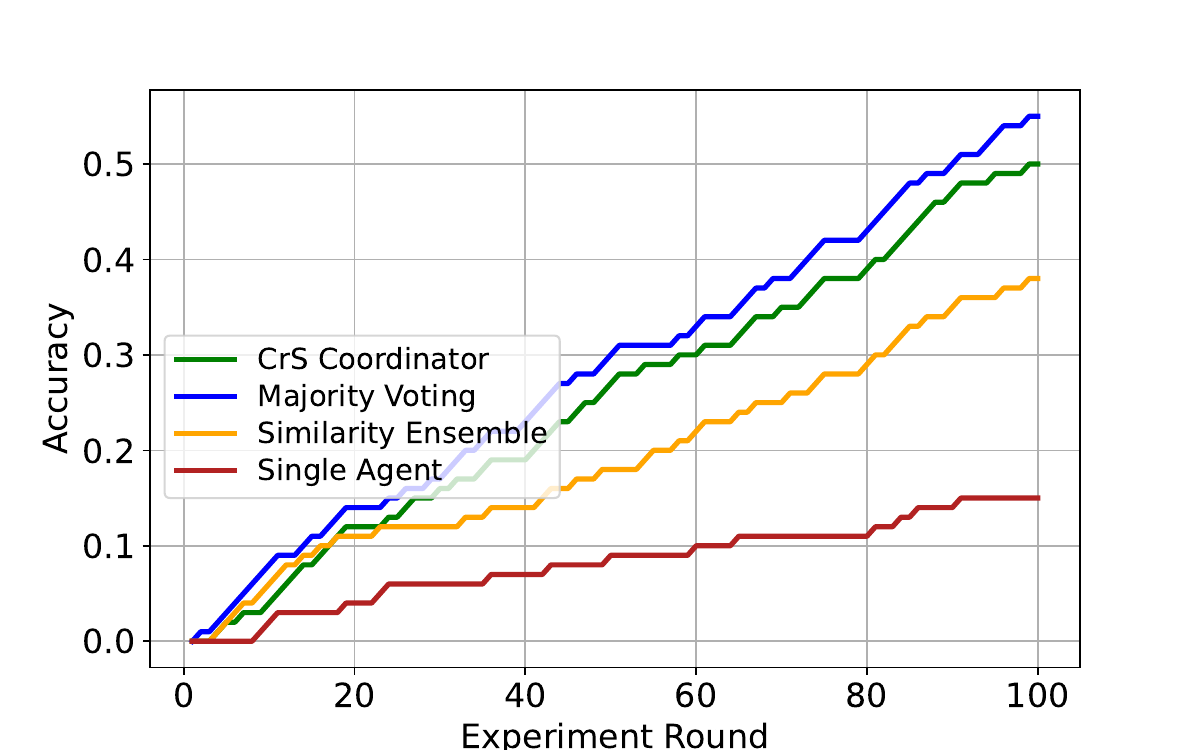}}
  \hfill
  \subcaptionbox{GSM8K-Mistral\label{fig:plotE}}[0.30\textwidth]{%
    \includegraphics[width=\linewidth]{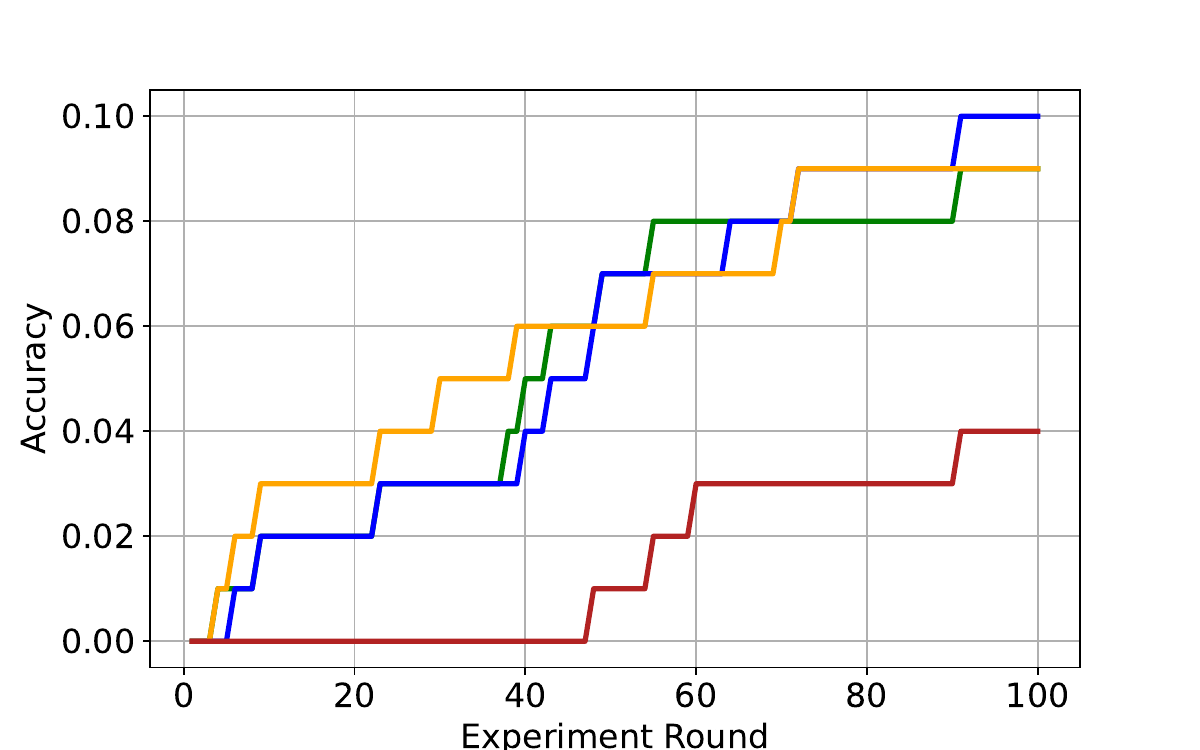}}
  \hfill
  \subcaptionbox{GSM8K-Qwen2.5\label{fig:plotF}}[0.30\textwidth]{%
    \includegraphics[width=\linewidth]{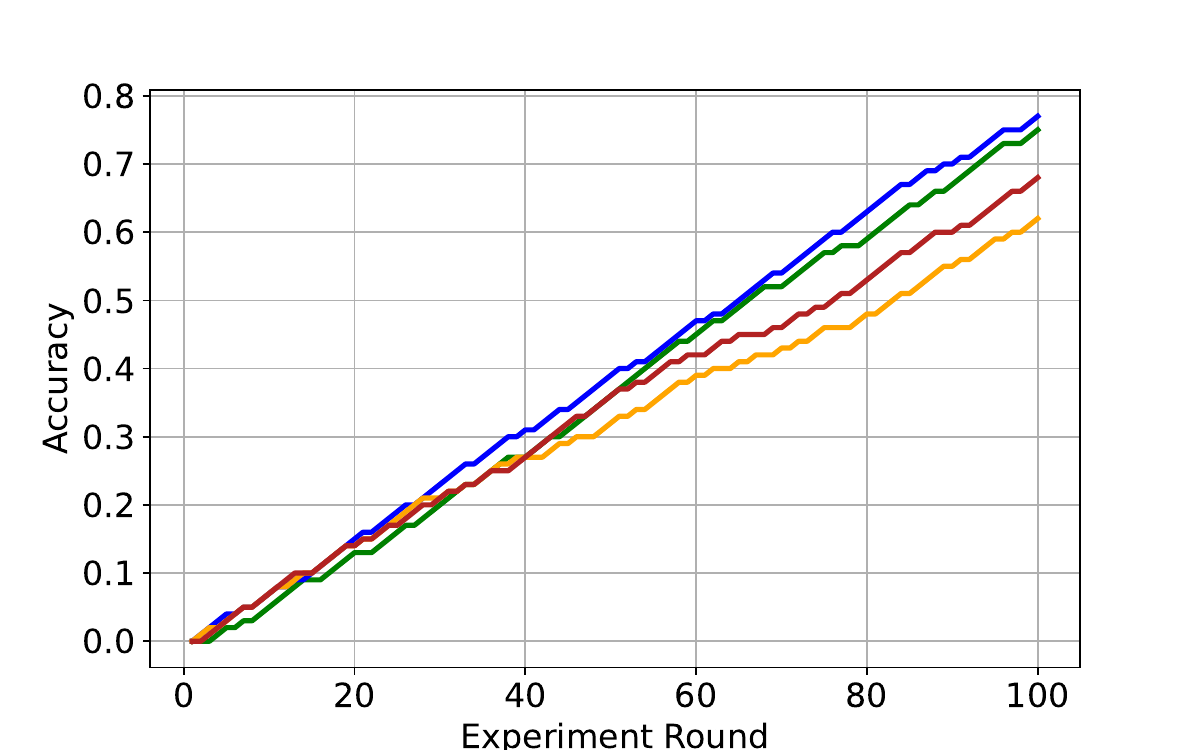}}

  \caption{Performance comparison of baseline methods versus CrS‑based coordinators.}
  \label{fig:sixplots}
  \vspace{-2mm}
\end{figure*}

 \begin{figure}[ht]
    \centering
    \begin{minipage}[b]{0.8\columnwidth}
    \includegraphics[width=\textwidth]{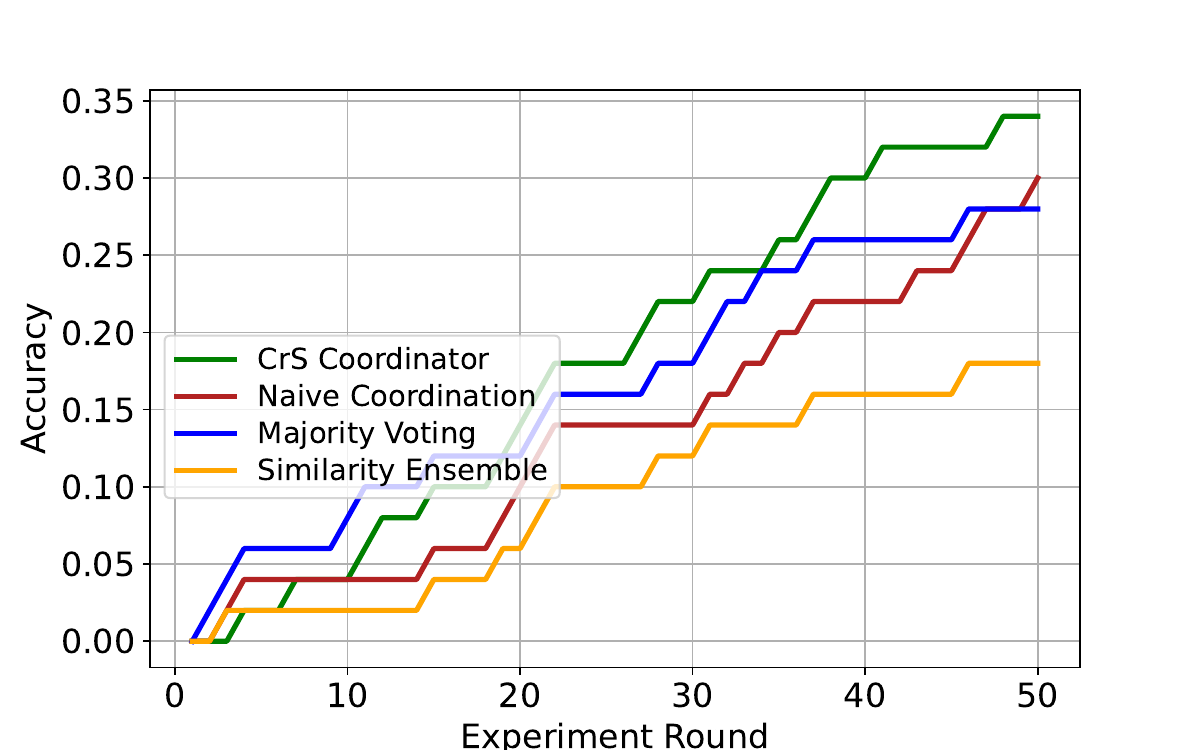}
    \caption{Baseline accuracy for a five‑agent chain (one faithful, four adversarial). The “CrS Coordinator” (green) curve reflects a CrS‑ordered chain, whereas all other methods use an unordered chain topology.}
    \label{fig:chain-baseline}
    \end{minipage}
    \vspace{-3mm}
\end{figure}

\subsubsection{Credibility Scores Drive Consistent Gains}
Across \textbf{all four benchmarks} (MMLU, GSM8K, MATH, ResearchQA) and for \textbf{every backbone} (LLaMA3.2, Mistral7B, Qwen2.57B), introducing our Credibility Score (CrS) raises accuracy by \emph{6--30 percentage points}. In high--noise settings such as GSM8K with three adversaries, CrS lifts LLaMA3.2 from \textbf{23\%$\rightarrow$42\%} in CrS-ordered chain and Qwen2.5 from \textbf{65\%$\rightarrow$75.5\%} in SIA. These patterns confirm that weighting agent opinions by empirically-measured reliability is a general mechanism for mitigating adversarial influence. 

Table~\ref{tab:SAA} presents results for Standalone Agent Architecture (SAA), which features no inter-agent interactions and utilizes a centroid-based aggregator inspired by~\cite{requal} as the coordinator. Our findings reveal consistent improvements in the number of correct responses. Specifically, in mathematical reasoning tasks such as GSM8K and MATH, the use of CrS coordination enhances the rate of fully correct responses $(r=1.0)$. This improvement occurs primarily by reducing the partially correct responses $(0.5 \leq r < 1.0)$, achieved through assigning higher weights to answers from more credible agents.

\subsubsection{Reasoning vs Multi-Choice Tasks}
We implement all three baseline models on an identical topology, utilizing the same agents to ensure consistency. Thus, the sole differentiating factor across these baselines is the coordination mechanism, allowing for a fair and precise comparison among models. Figure~\ref{fig:plotA},~\ref{fig:plotB} and ~\ref{fig:plotC}  illustrates that across 100 evaluated questions, the CrS coordinator consistently outperforms other baseline methods when confronted with a majority of adversarial agents. Interestingly, Majority Voting emerges as the second most effective coordination method after CrS. This result may initially seem counterintuitive, given that a majority of agents are adversarial and therefore expected to provide incorrect responses. However, as demonstrated in Table~\ref{tab:example-table-1}, adversaries occasionally alter their initial responses, eventually aligning with the correct solution. This phenomenon can be explained in two ways: 1) adversaries sometimes strategically shift their responses after misleading other agents to avoid revealing their adversarial nature; and 2) \textbf{adversaries can be influenced and persuaded by faithful agents}, prompting them to correct their earlier mistakes. Consequently, if at least one faithful agent consistently maintains the correct response, Majority Voting can yield accurate outcomes in specific scenarios. Nonetheless, these occasional successes are insufficient to prevent an overall decline in accuracy, reinforcing the superior robustness of the CrS coordinator against adversarial influence on MMLU-MS.
Figures~\ref{fig:plotD}, \ref{fig:plotE}, and \ref{fig:plotF} illustrate the performance of CrS on mathematical reasoning tasks using the GSM8K dataset. In these experiments, CrS achieves the second-best results, trailing behind Majority Voting. We attribute this performance gap to the intricate process of calculating Contribution Scores (CSc) in mathematical reasoning, where the complexity of reasoning significantly increases the likelihood of errors. These inaccuracies can corrupt the credibility score calculations and weighting mechanisms used by the CrS coordinator, occasionally resulting in the inadvertent prioritization of adversarial responses. This issue does not arise in Majority Voting. Nevertheless, despite these challenges, the CrS coordinator consistently outperforms Single Agent, Similarity Ensemble and naive coordination(Table~\ref{tab:progress-results}).

\begin{table}[t]            
  \small                    %
  \caption{Standalone Agent Architecture with LLaMA3.2(3B) agents. The table shows coordinator accuracy using credibility‑score (CrS) weights versus uniform weights across all tasks; numbers in parentheses indicate the resulting performance gap.}
  \label{tab:SAA}
  \begin{tabularx}{\columnwidth}{l|CC}  
    \hline
    \textbf{Dataset} & \textbf{\parbox[c]{2cm}{\raggedright Correct\\$(r=1.0)$}} & \textbf{\parbox[c]{2cm}{\raggedright Partially \\Correct$(0.5\le r<1.0)$}}  \\ \hline
    GSM8K & 57.6(\textcolor{darkgreen}{$+3.6\%$}) & 9.9(\textcolor{red}{$-1.6\%$}) \\
    MATH & 32.75(\textcolor{darkgreen}{$+5.75\%$}) & 13.3(\textcolor{red}{$-5.7\%$}) \\
    ResearchQA & 0.0 & 89.0(\textcolor{darkgreen}{$+2\%$}) \\
    MMLU-MS & 37.0(\textcolor{darkgreen}{$+2\%$})& - \\ \hline
  \end{tabularx}
  \vspace{-4mm}
\end{table}
\vspace{-1mm}
\subsubsection{Model Capacity Matters But Only With Coordination}
Small models (e.g., Mistral7B on MATH) suffer the steepest drops when exposed to adversaries: their multi-agent accuracy falls by up to \textbf{50\%} ($6/12$, Table~\ref{tab:progress-results}). CrS partially restores performance (\textasciitilde6pp gain), yet never reaches the ceiling attained by larger or instruction-tuned models. This suggests that \emph{coordination cannot fully compensate for insufficient backbone reasoning capacity}; future work might explore knowledge-distillation style training to narrow this gap.

\subsubsection{Judge-Computed CrS Imitates the Shapley Value }
\label{sec:exp:judge}

We illustrate the progression of CrS in Figures~\ref{fig:judge-weight} and \ref{Crs-Shapley}. Specifically, Figure~\ref{fig:judge-weight} presents the CrS evolution for Qwen2.5 agents on GSM8K—achieving the highest overall accuracy—and LLaMA3.2 agents on ResearchQA—demonstrating the greatest accuracy improvements, as detailed in Table~\ref{tab:progress-results}. The calculated CrS values effectively reflect agent credibility by appropriately down-weighting adversarial agents based on their contribution and reward metrics. Importantly, these CrS values closely approximate the Shapley value-based CrS used in the Standalone Agent Architecture (SAA), as evidenced by the consistent patterns in CrS progression and empirical outcomes. Further comparative results for both SAA and Stochastic Interaction Architecture (SIA) involving two agents (including one adversarial agent) are presented in Figures~\ref{crs-saa} and \ref{crs-sia} in Appendix~\ref{app:exp-ext}.


\subsubsection{Judge Alters the Outcome}

\begin{table}[h!]
\centering
\resizebox{0.9\columnwidth}{!}{%
\begin{tabular}{@{}lc|cc@{}}
\toprule
 & Pre-Communication & \multicolumn{2}{c}{Post-Communication} \\ 
&   & Chain & Random \\ 
\midrule
\textbf{CrS Coord.} & - & $0.16$& $0.12$ \\ 
Single Agent-LLaMa3.2(3B) & $0.32$ & $0.16$ & $0.16$\\ 
\bottomrule
\end{tabular}%
}

\caption{Comparison of accuracies before and after communication on a sample of 50 questions from HumanEval dataset.}
\label{tab:humaneval_comp}
\vspace{-2mm}
\end{table}

Replacing GPT-4o mini with a less capable judge, such as LLaMa3.2 (3B), leads to significant declines in accuracy—even when employing CrS—as erroneous evaluations of contribution scores (CSc) corrupt the credibility metrics essential for updating agent credibility. For instance, Qwen2.5 achieves the highest accuracy on GSM8K, as indicated in Table~\ref{tab:progress-results}, but using Llama3.2 (3B) as the evaluator decreases this accuracy by 54\%. This clearly demonstrates the critical dependence of CrS effectiveness on the evaluator's quality. A comparison between Figure~\ref{fig:llama-judge} in Appendix~\ref{app:exp-ext} and Figure~\ref{fig:judge-weight} further supports this conclusion.

Another issue arises when the judge is not capable of accurately evaluating the final response and providing a correct reward signal ($r$). This problem was particularly noticeable in our experiments on the HumanEval code completion benchmark using the GPT-4o mini judge (Table~\ref{tab:humaneval_comp}). These inaccurate evaluations, and the resulting miscalculations of Contribution Scores (CSc), significantly distort the Credibility Score (CrS) updates, ultimately undermining the overall effectiveness of the framework. While employing a more specialized and capable judge could reduce such inaccuracies, it also raises concerns about the practicality and necessity of the multi-agent configuration itself since directly assigning the task to a stronger evaluator might be more effective \footnote{A detailed analysis of the HumanEval results is provided in Appendix~\ref{app:exp-details}.}.

\subsubsection{Topology and Link Density}
\vspace{-2mm}
\begin{figure}[ht]
    \vspace{-2mm}
    \centering
    \begin{minipage}[b]{0.8\columnwidth}
    \includegraphics[width=\textwidth]{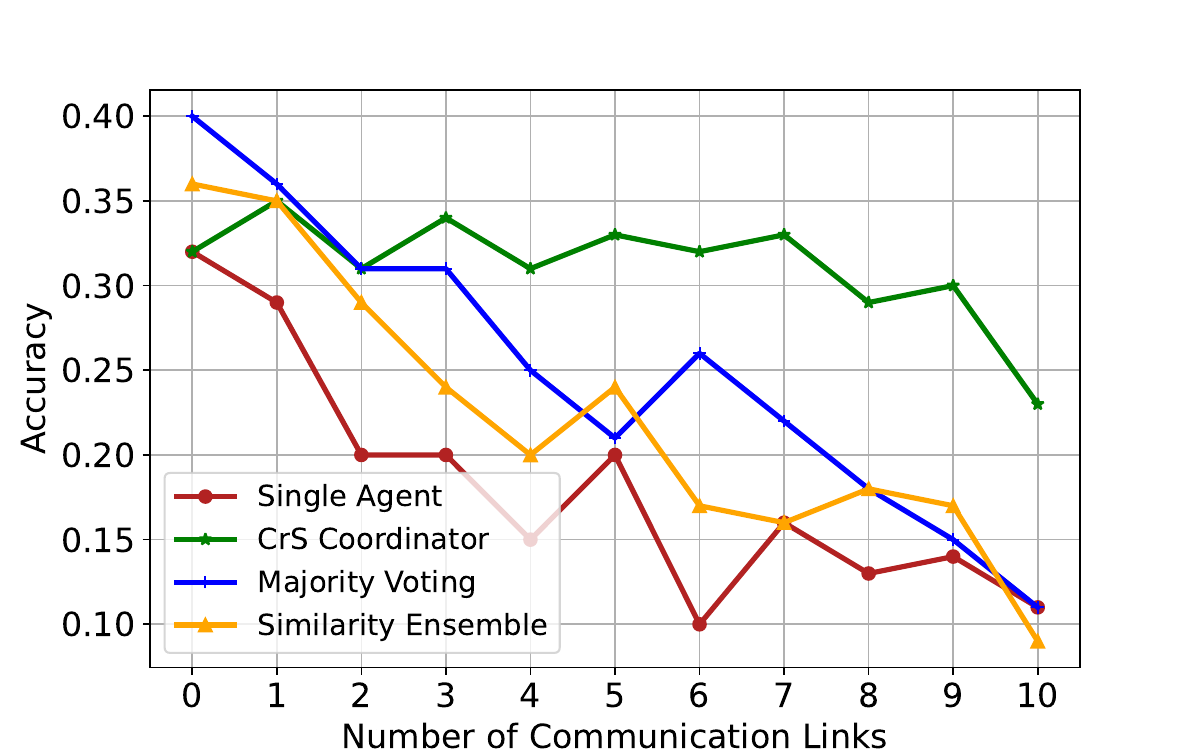}  
    \caption{Impact of the number of communication links on accuracy across baseline methods compared to the CrS coordination mechanism on MMLU-MS. }
    \label{fig:n_link}
    \end{minipage}
    \vspace{-2mm}
\end{figure}
Figure~\ref{fig:n_link} demonstrates that increasing the communication link count in SIA beyond six edges results in diminishing returns. Specifically, accuracy saturates at six links and notably \emph{decreases} when exceeding seven links, likely due to information overload. Conversely, increasing the link count extends the length of the communication history shared with the judge for computing the Contribution Score (CSc). This extension raises two primary concerns:
1) Activation of the token compressor becomes necessary to reduce token count to adhere to the judge's token limit requirements. This will increase the runtime.
2) If one round of token summarization is insufficient to meet these token requirements, subsequent rounds of compression may be triggered. Multiple rounds of compression risk losing information deemed non-essential by the compressor, ultimately affecting the accuracy and reliability of the contribution score.

Our empirical results indicate that a configuration with six links represents the optimal balance, effectively facilitating the study of adversarial impacts while minimizing the frequency of triggering more than one compression cycle. Figure~\ref{fig:n_link} also highlights the stability of the CrS coordinator with increased intra-group communication compared to other baseline methods, which exhibit a sharp decline in accuracy and significant negative impacts from additional communication rounds. The CrS coordinator's performance surpasses other aggregation methods by approximately 10 percentage points.
\vspace{-2mm}
\subsubsection{Adversary Proportion}\label{sec:exp-adversary}
\vspace{-1mm}
\begin{figure}[ht]
    \vspace{-4mm}
    \centering
    \begin{minipage}[b]{0.8\columnwidth}
    \includegraphics[width=\textwidth]{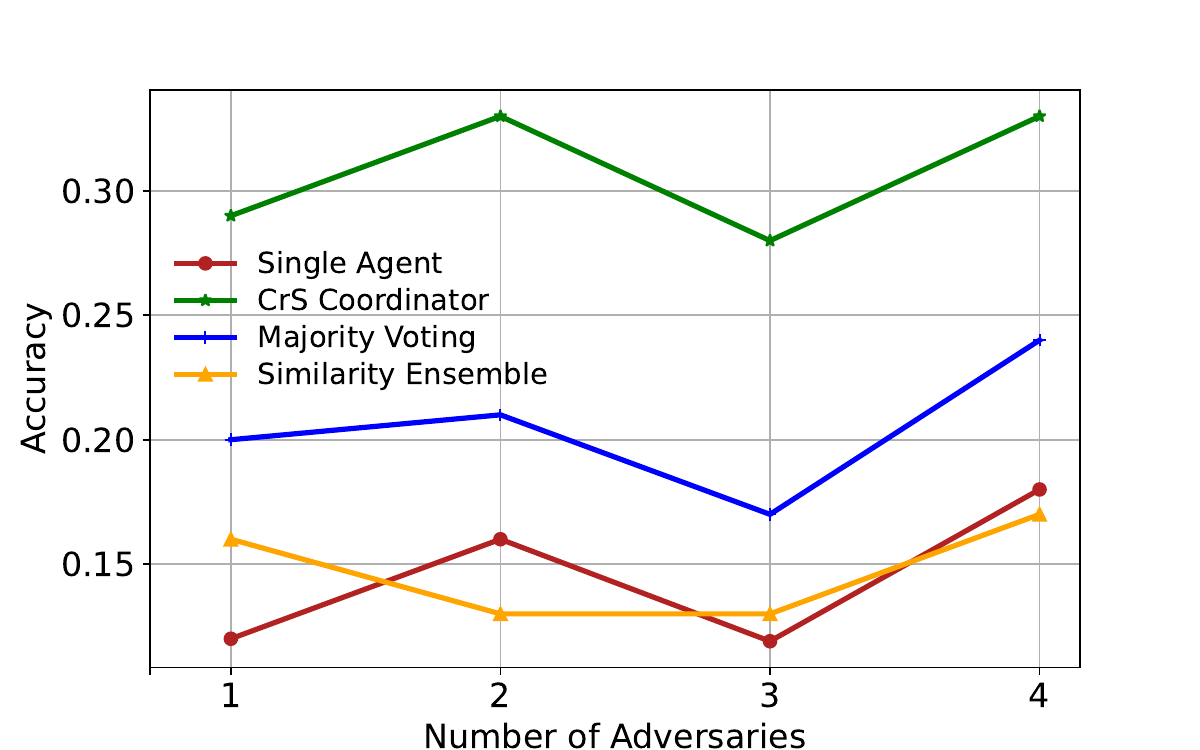}  
    \caption{Impact of adversarial agent count on accuracy across baseline methods compared to the CrS coordination mechanism on MMLU-MS.}
    \label{fig:n_adv}
    \end{minipage}
    \vspace{-3mm}
\end{figure}

Figure~\ref{fig:n_adv} demonstrates performance stability when employing CrS weighting: even with one to four adversaries present, accuracy consistently stays within a narrow range around 31\% ($\pm 2$ percentage points). In contrast, naive strategies experience significant fluctuations and never surpass 24\%. This stability indicates that reliability-based agent weighting effectively reduces sensitivity to adversary count, a promising outcome for scalability to larger and potentially noisier teams.

Figure~\ref{fig:chain-baseline} further supports this conclusion by demonstrating the superior performance of the CrS coordinator within a chain architecture, even under extreme adversarial conditions where 4 out of 5 agents are adversaries. These results validate our earlier findings in the Stochastic Interaction Architecture and suggest that the advantages of the CrS coordination mechanism extend reliably to structured communication topologies as well.



\section{Conclusion}\label{sec:conclusion}
In this paper, we introduced a general framework for building adversary-resistant multi-agent LLM systems using credibility scoring. By dynamically evaluating and weighting agents based on their contributions, our method enhances robustness against low-performing and adversarial agents, including in adversary-majority settings. This approach is adaptable to various team structures and task domains, offering a practical solution for securing multi-agent collaboration in LLM-based systems.
\section{Limitations}
\label{sec:limitations}

Our study advances multi-agent LLM coordination through Credibility Scores (CrS), yet several important limitations must be acknowledged.

\paragraph{Limited Evaluation Domains.}
Our evaluation focused exclusively on four benchmarks: MMLU, GSM8K, MATH, and ResearchQA. While these datasets collectively assess reasoning, coding, and factual question-answering capabilities, they do not encompass dialogue interactions, vision-language tasks, or real-time communication scenarios. Consequently, the generalizability of our findings to other contexts is limited.

\paragraph{Judge Dependence.}
The effectiveness of the CrS mechanism critically relies on the capabilities of the evaluator (judge). We observed significant performance degradation when employing weaker judges (e.g., LLaMA3.2 compared to GPT-4oMini). In such cases, Contribution Scores become noisy and lead to reduced accuracy (see Section~\ref{sec:exp:judge}). Future research could mitigate this sensitivity by developing self-calibrating judges or employing ensembles of judges.

\paragraph{Synthetic Adversaries.}
Our adversarial agents were explicitly instructed to exhibit adversarial behaviors and typically became easier to influence after multiple rounds of interaction. However, real-world adversaries, whether human actors or LLMs specifically fine-tuned for deceptive behaviors, may exhibit more sophisticated and unpredictable patterns. Such advanced adversaries could potentially evade detection or mitigation through CrS.

\paragraph{Computational and Cost Overheads.}
The computation of Shapley-like CrS scales quadratically with the number of agents involved, posing significant computational challenges. Each communication round necessitates two API calls to an external judge—one to evaluate the group's final response and another to review the interaction logs and compute Contribution Scores. These repeated calls incur substantial financial costs, limiting our ability to experiment with more powerful judges such as GPT-4o. This constraint particularly impacts tasks like HumanEval, where judge proficiency significantly influences reward calculation accuracy. Additionally, as the number of agents and communication links increases, interaction logs lengthen, triggering token-compression mechanisms. Such compression introduces additional latency and may result in the loss of critical context, further exacerbating evaluation inaccuracies. Exploring cost-effective approximations or more efficient evaluation techniques represents valuable avenues for future research.
\bibliography{ref}
\bibliographystyle{acl_natbib}
\appendix
\section*{APPENDIX}
\section{Related Work}\label{app:related}

As large language models (LLMs) continue to exhibit impressive capabilities in text comprehension \cite{llm-in-edu}, language generation, and reasoning \cite{reasoning-llm}, there is an increasing inclination to treat them as autonomous \underline{agents}, akin to humans. This perspective is reinforced by their ability to demonstrate human-like social behaviors that align with foundational theories in social psychology \cite{llm-human-agent}. However, despite these advancements, numerous studies \cite{llm-in-edu, long-text-qa, survey-llm-shortcomings, CoA} highlight persistent challenges in key areas, including mathematical reasoning, coding, and complex logical inference, as well as difficulties in processing long texts and generating extended narratives.


To overcome these limitations and improve factuality and reasoning, researchers have increasingly explored \underline{collaborative} problem-solving among multiple LLM agents rather than relying on a single model \cite{long-text-qa, survey-llm-shortcomings, MALLM-survey, MALLM-survey-2}. Similar to human teams that enhance their performance through collaboration, discussion, and iterative refinement, recent studies investigate whether LLMs can benefit from cooperative interactions. This paradigm shift leverages collective intelligence among LLM agents, allowing them to divide complex problems into manageable subtasks, particularly for more demanding and intricate problems. In these works, multiple LLM agents have been assembled to improve task performance through structured debate \cite{MA-LLM-debate, improving-fact-llm, div-think-llms} or ensemble methods \cite{more-agents}. 


Research in multi-agent LLM systems has yielded significant advancements, leading to the development of powerful frameworks such as CAMEL, AutoGen, and MetaGPT \cite{autogen, metagpt, camel}. These systems have demonstrated promising performance in crucial domains, including coding, mathematical problem-solving, and collaborative decision-making among multiple agents.


Despite these advancements, multi-agent LLM systems introduce inherent risks. If a subset of agents within the team is compromised—whether through poisoning attacks or adversarial intent—the collective output of the system can be corrupted. LLM agents are susceptible to persuasion, potentially leading them to reach incorrect consensus within the group. While previous studies \cite{psysafe, multiagent-debate-attack, MALLM-survey-2} have identified this vulnerability, existing solutions are primarily designed for specific, predefined architectures. 


This approach enhances prior multi-agent methods like the one by \citet{w_based_voting}, which used adversarial debate and credibility-weighted voting to reduce hallucinations. Instead of relying solely on inter-model disagreement, each LLM agent in this framework first undergoes internal self-refinement: tracking its own errors, measuring variance across multiple responses, and triggering self-reflection if thresholds are exceeded. Only after this process do agents engage in weighted voting, with conflicting outputs resolved through chain-of-thought comparisons. A final summarizing model then verifies consistency and coherence across agents. While this multi-phase design aims to improve robustness and factual accuracy, it implicitly assumes cooperative agents, making it vulnerable in adversarial settings. Moreover, the reliance on a summarization model that is stronger than the regular agents for final validation raises the question of why the task isn’t delegated to that model entirely.


To the best of our knowledge, {\em there is currently no general framework that enables users to design robust multi-agent systems resilient to adversarial influence while minimizing the impact of such attacks without the need to eliminate an agent}.


One approach, proposed by \cite{DyLAN}, introduces a query-based method to dynamically select the most influential agents within a multi-step feedforward network. However, this method relies on agents evaluating both themselves and their peers to assign \textit{Agent Importance Score}, making it particularly vulnerable in adversarial settings where malicious agents can manipulate the selection process and consensus within the group. 

In summary, existing literature proposes various coordination mechanisms—such as weight-based voting, expert specialization, and moderated debate—to improve robustness against adversarial agents, showing promising initial results~\cite{w_based_voting, div-think-llms}. 
However, no single solution effectively addresses all adversarial conditions; these mechanisms may still fail when adversaries form the majority or when the moderating model lacks significant superiority over adversarial agents.

\section{Incentives and Adversarially-behaving Agents}\label{app:behavior}

In multi-agent systems, the interplay between incentives and adversarial behavior significantly influences how agents interact and collectively function. Malicious agents pose a substantial risk by potentially undermining collective outcomes through tactics such as data or communication "poisoning." To mitigate these threats, robust defensive measures, including credibility or trust scores, are crucial for limiting the negative influence of adversarial agents. Carefully structured incentive mechanisms can either promote cooperation when agents share common objectives or effectively regulate the influence of self-interested agents with differing goals on the final outcome.

A multi-agent system requires mechanisms to assess reliability, reward trustworthy behavior, and penalize dishonest or consistently erroneous agents. This approach ensures that agents engaging in malicious or detrimental actions gradually lose their ability to influence collective decisions. Similar to human social dynamics, we propose that Large Language Model (LLM) agents also adopt distinct roles and vary in their levels of influence within a collaborative group.

To systematically evaluate an agent's significance in collaborative scenarios, we introduce the Contribution Score (\textit{CSc}). Inspired by the Shapley value—originally employed to measure the importance of individual features in linear regression tasks—the Contribution Score quantifies the impact each agent has on the group's overall performance \cite{shap}. While this metric effectively captures an agent's overall influence within the group, it does not inherently differentiate between positive and negative contributions. Consequently, an agent can attain a high Contribution Score despite disseminating adversarial or misleading information, adversely affecting the group's final outcomes. 
To effectively address this challenge, we introduce the \textit{Credibility Score} (\textit{CrS}), which is initially assigned uniformly across all agents and dynamically evolves throughout successive iterations, serving as an agent profiling mechanism. 

\section{Coordination Mechanisms}\label{app:coord}

In multi-agent systems, coordination mechanisms determine how individual agents' outputs are integrated. 
A critical component of coordination is the aggregation approach, which may include techniques such as majority voting, weighted averaging, or the utilization of specialized coordinator agents responsible for synthesizing multiple agent solutions into a cohesive outcome. In the following we briefly discuss each method.

\paragraph{Majority Voting}
 Each LLM agent produces an answer, and the ensemble selects the option most frequently proposed. In both self‑consistency decoding where multiple independent samples from a single model—and true multi‑model ensembles, majority vote reliably boosts accuracy because uncorrelated errors are out‑voted by repeated correct answers~\cite{self-consistency}. Its effectiveness scales with the number of agents, allowing a group of small LLMs to rival a single larger model~\cite{more-agents}. However, previous studies indicate that when adversarial or malicious behaviors are present in at least half ($N/2$) of the agents in a group of size $N$, traditional aggregation methods like majority voting become considerably less effective \cite{more-agents, multiagent-debate-attack}.

\paragraph{Weighted Averaging}
A generalization of majority vote assigns each agent a reliability weight, and the ensemble picks the answer backed by the highest total weight from all agents. Systems such as ReConcile~\cite{reconcile} and Boosted Prompt Ensembles~\cite{boosted-prompt} show that emphasizing historically accurate agents achieves higher overall accuracy and partial robustness to noisy or malicious peers. However, performance hinges on correct weight estimation; if adversaries obtain high weights, they can still dominate the ensemble.

\paragraph{Similarity-Based Ensemble}

Rather than relying on explicit voting, similarity-based ensemble methods select the response that is most semantically aligned with all others, assuming that the correct answer will form the tightest consensus cluster. Smoothie \cite{smoothie} and Agent-Forest \cite{more-agents} operationalize this by embedding candidate answers into a vector space and choosing the one with the lowest average distance to its peers, achieving strong performance without the need for supervised weights. These approaches naturally filter out outliers but remain vulnerable to coordinated adversarial agents that produce highly similar incorrect responses.

\paragraph{Centroid-based Aggregation} 
Ebrahimi et al. \cite{requal} extend similarity-based ensemble by combining weighted averaging with similarity-based selection. They propose a Monte Carlo-based strategy that selects the response closest to a weighted centroid of all answers, where the weights $w_i$ reflect the agents' reliability. 
The centroid vector, $\vec{x}^+$, is computed as a weighted average of the generated responses in the embedding space, i.e., $\vec{x}^+ = \frac{1}{|R|}\sum_{i=1}^{|R|} w_i.\vec{v}(x_i)$. Then, the final answer is identified as
\begin{equation}
x^{\star}=\arg\min_{x\in R} d(\vec{v}_x,\vec{x}^+)
\end{equation}
where $d(\cdot,\cdot)$ is the cosine distance between embeddings.
In our work, we adopt this aggregation method in a no inter-agent communication setting, using credibility scores as weights to guide the centroid-based coordination process.

\paragraph{LLM-based Coordination}

Recent works suggest that an llm-based \textbf{coordinator agent} is an effective aggregation mechanism for multi-agent systems. \cite{MA-LLM-debate} show that letting the agent debate before a coordinator renders the final verdictcan improve the overall accuracy. Yet, they warn that malicious participants may still steer the group toward suboptimal answers. Subsequent studies explore two types of task distribution paradigms: i) \textbf{redundant solving}, where the agents tackle same prompt to gain robustness through majority consensus and ii) divide-and-conquer where a complex task is broken into subtasks whose answers must be carefully integrated. In both setting a coordinator (or a manager) LLM synthesises the individual responses into a coherent final answer, mitigating inconsistencies and guarding local errors.
This manager-style coordination has been adopted in recent multi-agent LLM frameworks such as \cite{CoA} and \cite{MoA}, which report higher overall accuracy and improved resilience to adversarial or noisy agents compared with uncoordinated ensembles.

\section{Reward Calculation}\label{app:reward}

Evaluation and feedback ensure that agents’ contributions are measured against some reliable standard. Often, a ground truth or external judge is used to compare the collective, final solution with a correct reference or quality metric. This judge can be an oracle, a human evaluator, or a specialized LLM that scores how accurate or useful each solution is \cite{researchQA}. The resulting reward/penalty can then guide learning, credibility score updates, ultimately improving the system’s performance over time. 

We propose a comprehensive framework suitable for a variety of scenarios, emphasizing preventive measures to penalize adversarial behavior and facilitating informed aggregation to improve decision-making reliability. Our framework comprises three key components: 1) a team of agents organized into diverse topologies to accommodate multiple modes of multi-agent collaboration, 2) an evaluation mechanism designed to objectively assess the performance of individual agents, and 3) a coordination mechanism that systematically integrates agent responses. Furthermore, we introduce two critical metrics—the Credibility Score (Src) and the Contribution Score—to effectively measure each agent's reliability and contribution. These components are designed flexibly, allowing our method to adapt seamlessly to any collaboration graph topology and coordination strategy.
\section{Experiments Setting Details}\label{app:exp-details}

\paragraph{Backbone Models.} Although powerful models such as GPT-4 exhibit notable robustness to adversarial interference, smaller and less sophisticated models remain highly vulnerable, experiencing significant accuracy drop under adversarial conditions. To effectively assess the robustness and efficiency of our proposed framework, we select lightweight open-source models as the backbone for both individual agents and the coordinator. Lightweight models offer the advantage of resource-efficient loading and execution, thus ensuring scalability and practicality in multi-agent settings. Specifically, we employ LLaMA 3.2 (3B)~\cite{llama3.2}, Mistral (7B)~\cite{mistral7b}, and Qwen2.5 (7B)~\cite{qwen2.5} as our backbone models.
Moreover, we utilize GPT-4o mini~\cite{gpt4o-mini} as an external judge to assess the quality and correctness of the final responses generated by the multi-agent team.

\paragraph{Datasets.} We evaluate the effectiveness of our proposed framework across five benchmark datasets: MMLU~\cite{MMLU}, MATH~\cite{math}, GSM8K~\cite{gsm8k}, HumanEval~\cite{humaneval}, and Research Questions~\cite{researchQA}. Specifically, we use high school mathematics and statistics questions from the MMLU dataset, which are referred to as MMLU-MS, to assess the performance of the model in multiple-choice question answering. The MATH and GSM8K datasets are employed to evaluate mathematical reasoning capabilities, while HumanEval is used to assess coding proficiency. The Researchy Questions dataset consists of non-factoid questions derived from real-world search engine queries, characterized by their subjective nature and absence of a singularly correct answer. In this context, a human judge or an external judge must carefully evaluate agent responses, determining correctness based on provided instructions and contextual information. 

\subsection{Collaboration Setup}

Our primary experiments involve a team of five agents, comprising two faithful agents and three adversarial agents explicitly instructed to introduce subtle inaccuracies in their responses without revealing their adversarial nature. We employ consistent prompts across various tasks, adapting only the task-specific details. Our analysis primarily explores two main communication structures: a Stochastic Interaction  Architecture (SIA) , Standalone Agent Architecture (SAA) and a Credibility-ordered Chain.

\subsubsection{Standalone Agent Architecture (SAA)}

In SAA every agent receives the same question $Q$ and produces an answer \emph{without any communication}.  
The resulting communication graph is edgeless: $G=(\mathcal{A},\emptyset)$.
We aggregate the set of agent answers $R=\{x_1,\dots,x_{|R|}\}$ using the centroid-based ensemble method of \cite{requal}.  
Let $v(x)$ be the embedding of answer $x$ and let $w_i\propto\operatorname{CrS}^{(i)}$ be the credibility weight of agent~$i$.  
The credibility‑weighted centroid is  
\[
\mathbf{v}_c=\frac{1}{|R|}\sum_{i=1}^{|R|} w_i\,\mathbf{v}(x_i),
\]  
and the final answer is the one whose embedding is closest (cosine distance $d$) to that centroid:  
\[
x^\star=\arg\min_{x\in R} d\!\bigl(\mathbf{v}(x),\mathbf{v}_c\bigr).
\]  
Finally, we calculate each agent’s Contribution Score (CSc)—derived from the Shapley value as described in\S\ref{sec:tech-details-Csc}—and, from these, obtain the Credibility Scores (CrS) for the entire set of responses.

\subsubsection{Stochastic Interaction Architecture (SIA)}
SIA adds a sparse, random communication graph $G_t$ that is resampled for every query.  
For each question we draw $m$ undirected edges from the $\binom{N}{2}$ possible pairs with replacment ($N= 5, m=6$ in our experiments), typically creating six links.  
Connected agents exchange their current answers and may revise them, producing diverse topologies such as trees, rings, and other sparse structures—while avoiding full information saturation that would otherwise aid adversaries.

\subsubsection{Credibility-ordered Chain}
To further test our hypothesis within a specific, stable structure, we introduce the chain-based architecture. 
In the chain architecture agents are sorted in descending order of their credibility score in the beginning of the experiment. Communication in this structure only occurs between adjacent agents. 
Positioning the most reliable agents earlier in the chain limits the influence of adversaries further down the chain. Although the communication pattern remains fixed, CrS values continue to be updated throughout the interactions within the chain.


\subsection{Why Three Architectures?}
SAA provides a lower bound on performance—no interactions means no adversarial persuasion—while SIA explores the hard regime where adversaries may form majorities and hijack discussions by persuading other agents. The credibility‑ordered chain tests our hypothesis in a stable yet asymmetric structure. Experiments in \S\ref{sec:experiments} demonstrate that CSc/CrS significantly improve robustness across all three settings.

\section{Extended Experiment Results}\label{app:exp-ext}

\subsection{Judge Alters the Outcome}

The effectiveness of the judge is highly task-dependent. To illustrate this, we present results on two different benchmarks: HumanEval for code completion, and GSM8K for mathematical reasoning.

On the HumanEval benchmark, using GPT-4o mini as the judge proves problematic. In this task, the judge receives a reference solution, a set of test functions, and the final response generated by the CrS coordinator. However, it often fails to correctly determine whether the generated code is functionally correct. This results in numerous cases where incorrect solutions are mistakenly rewarded with a score of 1, severely distorting the Contribution Scores (CSc) and, consequently, the Credibility Scores (CrS). As shown in Table~\ref{tab:humaneval_comp}, these misjudgments ultimately hurt overall system accuracy.

\begin{figure}[ht]
    \includegraphics[width=0.5\textwidth]{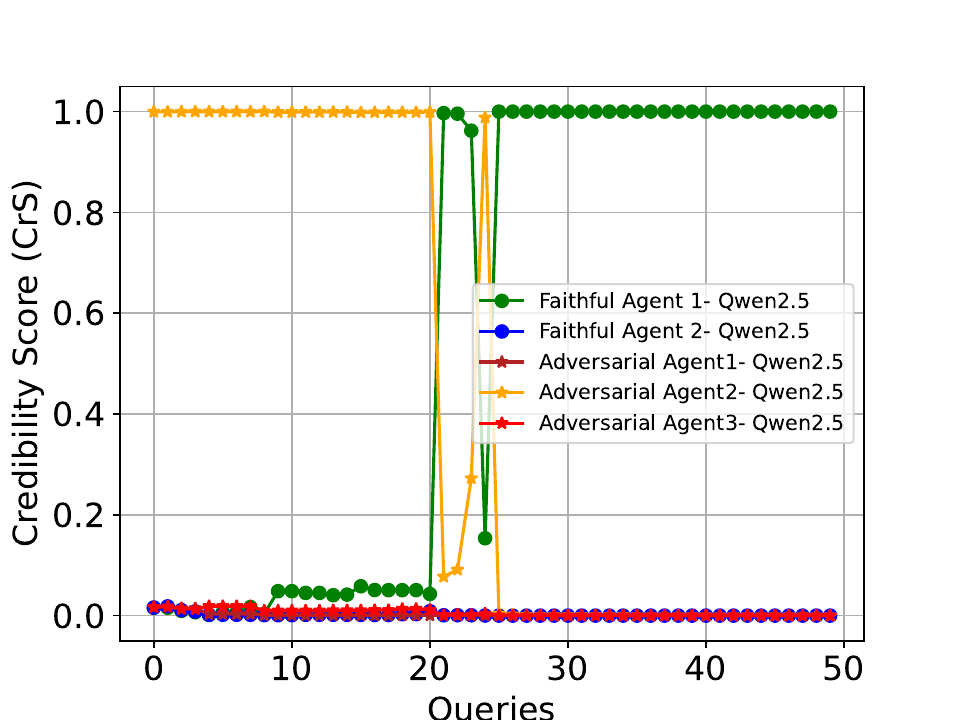}
    \caption{CrS evolution with a LLaMA‑3.2(3B) judge supervising five Qwen2.5(7B) agents on GSM8K—directly comparable to Figure\ref{fig:judge-weight-a}. }
    \label{fig:llama-judge}
\end{figure}

For mathematical reasoning questions from GSM8K, we observe a different failure mode when using a weaker judge. Figure~\ref{fig:llama-judge} shows the CrS trajectories for five agents—two faithful and three adversarial—when LLaMA 3.2's is used as the evaluator. Compared to the more stable CrS patterns seen with GPT-4o mini (Figure~\ref{fig:judge-weight}), the scores here fluctuate significantly. This instability stems from LLaMA 3.2's tendency to produce malformed outputs or to incorrectly assess agent contributions—such as returning a two-element array (e.g., $[0.2, 0.8]$) in a five-agent setting—indicating its limited ability to follow contribution-scoring instructions accurately.

\definecolor{softyellow}{RGB}{255, 249, 196}
\definecolor{softblue}{RGB}{226, 240, 255}

\begin{table*}[h!]
\centering
\small
\setlength{\tabcolsep}{4.5pt}
\renewcommand{\arraystretch}{1.1}
\begin{tabular}{lccccccp{2.8cm}c}
\toprule
\textbf{Agent} & L1 & L2 & L3 & L4 & L5 & L6 & \textbf{CrS (curr→fut.)} & \textbf{CSc} \\
\midrule
Agent 1: B & \cellcolor{softyellow}B & B & B & \cellcolor{softyellow}X & X & X & $0.4593 \rightarrow 0.4617$ & 0.15 \\
Agent 2: B & B & \cellcolor{softyellow}B & \cellcolor{softyellow}D & D & \cellcolor{softyellow}C & C & $0.4143 \rightarrow 0.4143$ & 0.20 \\
Agent 3: B & B & B & B & B & \cellcolor{softyellow}B & \cellcolor{softyellow}D & $0.4224 \rightarrow 0.4225$ & 0.20 \\
Agent 4: B & B & \cellcolor{softyellow}C & \cellcolor{softyellow}X & \cellcolor{softyellow}C & C & \cellcolor{softyellow}D & $0.4711 \rightarrow 0.4688$ & 0.25 \\
Agent 5: C & \cellcolor{softyellow}C & C & C & C & C & C & $0.4655 \rightarrow 0.4656$ & 0.20 \\
\midrule
\rowcolor{softblue}
\multicolumn{7}{r}{\textbf{Final Answer}} & C & \\
\rowcolor{softblue}
\multicolumn{7}{r}{\textbf{Correct Answer}} & D & \\
\rowcolor{softblue}
\multicolumn{7}{r}{\textbf{Reward}} & -1 & \\
\bottomrule
\end{tabular}
\caption{Compact illustration of agent response dynamics and credibility updates. Yellow cells indicate response changes.}
\label{tab:example-table-1}
\end{table*}

\definecolor{softyellow}{RGB}{255, 249, 196}
\definecolor{softblue}{RGB}{226, 240, 255}

\begin{table*}[h!]
\centering
\small
\setlength{\tabcolsep}{4.5pt}
\renewcommand{\arraystretch}{1.1}
\begin{tabular}{lccccccp{2.8cm}c}
\toprule
\textbf{Agent} & L1 & L2 & L3 & L4 & L5 & L6 & \textbf{CrS (→)} & \textbf{CSc} \\
\midrule
Agent 1: D & \cellcolor{softyellow}B & B & B & B & B & \cellcolor{softyellow}A & $0.4841 \rightarrow 0.4940$ & 0.00 \\
Agent 2: B & \cellcolor{softyellow}X & X & \cellcolor{softyellow}A & A & A & A & $0.3613 \rightarrow 0.3686$ & 0.00 \\
Agent 3: B & B & B & B & \cellcolor{softyellow}X & X & X & $0.4034 \rightarrow 0.3951$ & 0.40 \\
Agent 4: B & B & \cellcolor{softyellow}A & \cellcolor{softyellow}A & \cellcolor{softyellow}C & \cellcolor{softyellow}A & \cellcolor{softyellow}C & $0.4561 \rightarrow 0.4468$ & 0.40 \\
Agent 5: C & C & \cellcolor{softyellow}C & C & C & \cellcolor{softyellow}C & C & $0.5139 \rightarrow 0.5139$ & 0.20 \\
\midrule
\rowcolor{softblue}
\multicolumn{7}{r}{\textbf{Final Answer}} & C & \\
\rowcolor{softblue}
\multicolumn{7}{r}{\textbf{Correct Answer}} & B & \\
\rowcolor{softblue}
\multicolumn{7}{r}{\textbf{Reward}} & -1 & \\
\bottomrule
\end{tabular}
\vspace{0.5em}
\caption{Illustrative example from MMLU with two faithful agents. Although the final response was incorrect, these agents were not penalized— the judge identified adversarial influence from Agent 4 based on the communication history.}
\label{tab:highlighted_table}
\end{table*}

\begin{figure*}[t]              
  \centering
  \begin{subfigure}[b]{0.46\textwidth}
    \centering
    \includegraphics[width=\linewidth]{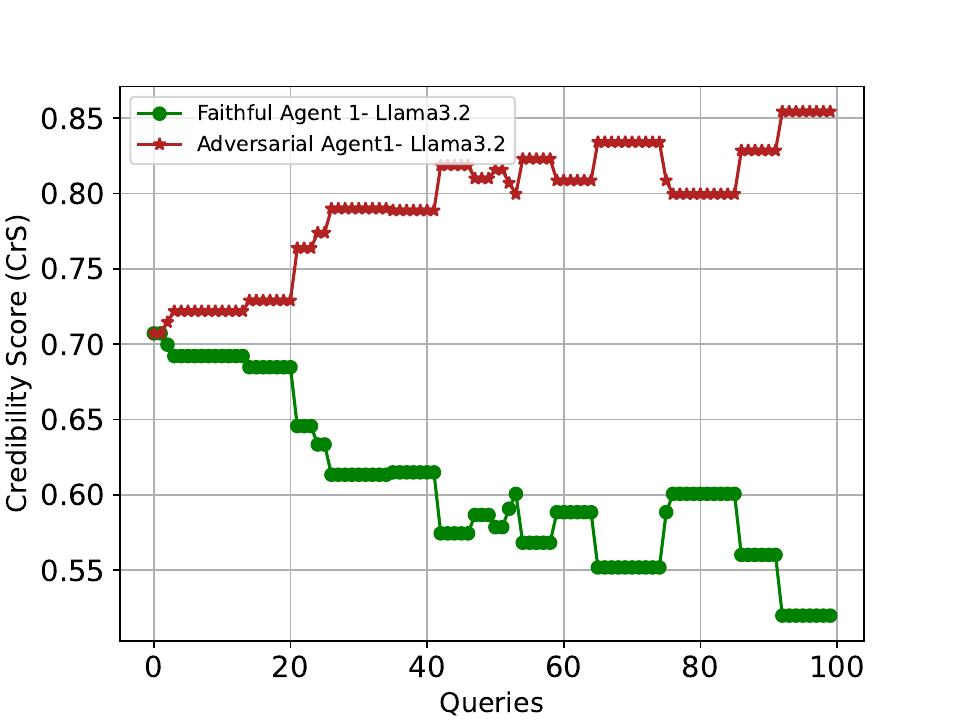}
    \caption{After exchanging messages, Each agent outputs a revised solution, and an identical LLaMA‑3.2 coordinator produces the final response using CrS‑weighted aggregation of their answers.}
    \label{crs-saa}
  \end{subfigure}%
  \hfill
  \begin{subfigure}[b]{0.46\textwidth}
    \centering
    \includegraphics[width=\linewidth]{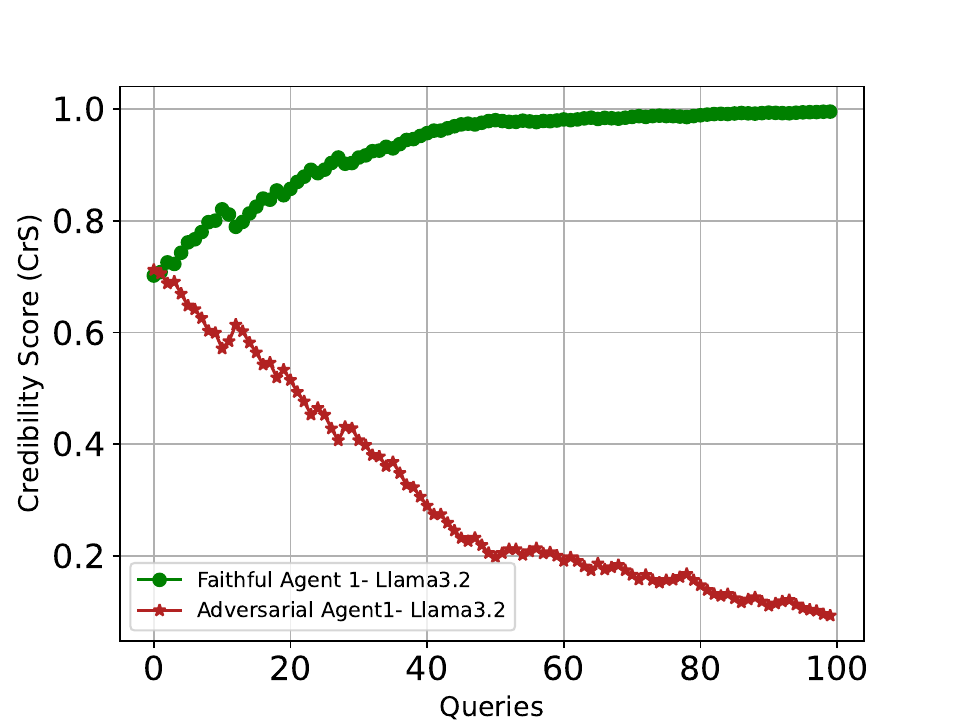}
    \caption{The agents have no inter‑agent communication (SAA). Each agent generates a candidate answer, and the coordination mechanism selects the answer nearest to their CrS‑weighted centroid.}
    \label{crs-sia}
  \end{subfigure}

  \vspace{0.5em}  
  \caption{CrS evolution for two independent LLaMA‑3.2 (3B).}
  \label{Crs-Shapley}
\end{figure*}


\end{document}